\documentclass[twocolumn,amsmath,letterpaper]{revtex4}
\usepackage{graphicx}

\newcommand\eg{\textit{e.g.}}
\newcommand\ie{\textit{i.e.}}
\newcommand\lhs{\textit{l.h.s.}}
\newcommand\rhs{\textit{r.h.s.}}
\newcommand\rms{\textit{r.m.s.}}
\newcommand\mrms{\mathrm{rms}}
\renewcommand\d{\mathrm{d}}
\newcommand\e{\mathrm{e}}
\newcommand\im{\mathrm{i}}
\newcommand\de[2]{\frac{\partial #1}{\partial #2}}
\renewcommand\vec[1]{\mathbf{#1}}
\newcommand\mat[1]{\mathbf{#1}}
\renewcommand\Im{\,\mathrm{Im}}
\renewcommand\Re{\,\mathrm{Re}}

\begin{document}
\title{The role of microscopic fluctuations in transition prediction}
\thanks{submitted for publication to \textit{Physics of Fluids}}
\author{Paolo Luchini}
\email{luchini@unisa.it}
\affiliation{DIMEC, Universit\`a di Salerno, Italy}

\begin{abstract}
The commonly accepted description of transition to turbulence in shear flows
requires the presence of an external source of disturbances that get amplified
by an essentially linear mechanism up to the point where breakdown to turbulence
occurs. Microscopic fluctuations are shown here to provide just the right amount
of initial disturbances to match the predictions of linear stability theory. 
\end{abstract}

\maketitle

\section{Introduction}
Things as different as Brownian motion and the input noise of an electronic amplifier (and Kirchhoff's law of thermal radiation, although not generally mentioned in this connection) have their common explanation in the fluctuation-dissipation theorem, which stipulates that every macroscopic dissipation phenomenon is accompanied by a random forcing induced
by the underlying microscopic mechanics, with a spectrum that is a universal function of temperature. In
fluid dynamics the appropriate forcing, in the form of a random stress
tensor, was added to the Navier-Stokes equations by Landau \& Lifschitz
\cite{LL} and further discussed by Fox \& Uhlenbeck \cite{FU}.

In almost the same years, the theory of linear instabilities of the
Navier-Stokes equations gradually evolved into an engineering tool for
transition prediction. Being able to determine the position of
transition from laminar to turbulent flow has a large practical impact on the
design of aircraft and many other man-made artifacts; at the same time it is a
problem that still defeats a complete solution, in spite of universal agreement that the Navier-Stokes
equations constitute an adequate physical model. It is significant that, in an often quoted sentence, Richard Feynman \cite{Feyn} chose a transitional flow to produce a didactic example of the inscrutability of differential equations: \textit{The next great era of awakening of human intellect may well produce a method of understanding the qualitative content of equations. Today we cannot. Today we cannot see that the water flow equations contain such things as the barber pole structure of turbulence that one sees between rotating cylinders}.

The starting point of numerical transition prediction was the (rather inexpressively named, but so known ever since) $\e^N$ method of Smith \& Gamberoni \cite{SG} and Van Ingen \cite{vI}. The crux of it was the realization that shear flows behave as amplifiers
(in a more recent language, they support \textit{convective} instabilities),
accepting small disturbances of various origins at some upstream location and
amplifying them downstream to the point where transition occurs. The original
version of the method only computes the linear-amplification factor
$\e^N$ of
Tollmien-Schlichting waves, from one of a few alternative quasi-parallel approximations of the Navier-Stokes equations, and uses the value of the exponent
$N$ as a transition criterion \footnote{This is in fact the amplification measured in \textit{neper}, albeit the neper as a measuring unit is not
generally mentioned in the transition-prediction literature.}, experiments having established that transition occurs at the location where $N\approx 9\div 10$ under the smallest environmental disturbances, or earlier otherwise.

In an effort to improve upon the $e^N$ method many researchers have focussed on the \textit{receptivity}
problem, \ie\ the coupling of external disturbances to the boundary layer that
precedes the amplification proper. For a review see, \eg, Saric \cite{Saric}. Different types of disturbances have been studied in a relatively large literature: acoustic waves, freestream
vorticity, wall roughness and the presence of a leading edge are all considered to be sources of the boundary-layer
instability mode that, once amplified, triggers transition. (Nonlinear mixing of at least two of these sources is generally necessary to produce the right phase speed.) However, a
stumbling block to the application of receptivity theory to transition
prediction has been the difficulty of specifying realistic amplitudes and
spectra for all these external disturbances. For this reason, most if not all
engineering computations are still today performed by the $\e^N$ method without any receptivity accounted for.

Despite microscopic fluctuations having been given attention by authors who pursue a general theory of turbulence (\eg\ \cite{Ruelle}) and, recently, in DNS of Couette flow \cite{Oblk}, surprisingly they have never been considered as candidate disturbance sources in the context of transition prediction. Yet the $\e^N$ method views the boundary layer as an amplifier; every amplifier has noise and the lower bound of possible noises is thermal noise, without which the second law of thermodynamics would be overthrown. Therefore it appears to be of both
theoretical and practical interest to calculate the receptivity of
boundary-layer instability modes to thermal noise and compare its effects to
$\e^N$ predictions. So this paper purports to do.

\section{Basic formalism of stochastically driven linear systems}
We briefly summarize the theory of stochastically driven linear systems so to establish a notation for the
required formulae. For a forced finite-dimensional linear system with state
vector $\vec{f}$ ruled by the evolution equation
\begin{equation}
\frac{\d \vec{f}}{\d t} = \mat{A}\, \vec{f} + \vec{g}(t)
\label{fds}
\end{equation}
we can define eigenvalues $\sigma^{(i)}$, and direct $\vec{u}^{(i)}$ and adjoint
$\vec{v}^{(i)}$ modes (respectively right and left eigenvectors of the evolution matrix $\mat{A}$). The Laplace transform of eq.(\ref{fds}) allows us to write its solution as
\begin{equation}
\vec{f} = \left(\sigma - \mat{A}\right)^{-1}\vec{g} = \sum_{i=1}^n
\frac{\vec{u}^{(i)}\;\vec{v}^{(i)}\cdot \vec{g}}{\sigma - \sigma^{(i)}}\, ,
\label{matexp}
\end{equation}
where the expansion of a matrix (with no multiple eigenvalues) in terms of its
eigenvectors (normalized so that $\vec{v}^{(i)}\cdot\vec{u}^{(i)}=1$) has been
used. In the time domain, the leading 
eigenvalue $\sigma^{(1)}$ (the one with the largest real part) will dominate for large enough time. Thus, for sufficiently large $t$,
\begin{equation}
\vec{f}(t) \simeq \vec{u}^{(1)} \int_0^t\e^{\sigma^{(1)}(t-\tau)}\vec{v}^{(1)}\cdot
\vec{g}(\tau) \d\tau\, .
\label{ltb}
\end{equation}
If now $\vec{g}$(t) is a statistically steady white noise with autocorrelation
matrix $\left<\vec{g}^*(t)\vec{g}(t')\right> = \mat{G}\,\delta(t-t')$ \footnote{a $*$ superscript denoting complex conjugation}, we shall have
\begin{multline}
\left<\left|\vec{f}(t)\right|^2\right> = \left|\vec{u}^{(1)}\right|^2
\int_0^t\int_0^{t}\e^{\sigma^{(1)*}(t-\tau)+\sigma^{(1)}(t-\tau')}\\
\times\vec{v}^{(1)*}\cdot\mat{G}\cdot\vec{v}^{(1)} \delta(\tau-\tau') \d\tau' \d\tau
=\\
= \left|\vec{u}^{(1)}\right|^2\; \vec{v}^{(1)*}\cdot\mat{G}\cdot\vec{v}^{(1)}\;
\frac{\e^{2\Re(\sigma^{(1)})t}-1}{2\Re(\sigma^{(1)})}\, .
\label{meansq}
\end{multline}
Equation (\ref{meansq}) is the main result we shall want to extend to
fluid-dynamic equations in the following. We note that in order to determine the
mean square value of $\vec{f}$ at time $t$ there is no need to specify the
probability distribution of $\vec{g}$, although in the case of thermal noise this generally is gaussian.

\section{Fluid-dynamic equations}
We can now apply a similar formalism to the infinite-dimensional system
governed by the partial differential equations
of fluid dynamics. We shall limit our attention to incompressible fluids with constant properties
in a parallel flow. The linearized Navier-Stokes equations with an added random stress term can be written as 
\begin{subequations}
\begin{equation}
\de{u_i}{x_i} = 0\, ,
\label{cont}
\end{equation}
\begin{multline}
\rho\left(\de{u_j}{t} + U\de{u_j}{x_1} + \delta_{1j}\frac{dU}{dx_2}u_2\right) +\de{p}{x_j} -\mu \de{^2u_j}{x_i\partial x_i} =\\
= \de{s_{ij}}{x_i} \, .
\label{mom}
\end{multline}
\label{physdom}
\label{transformed}
\end{subequations}
Here $u_1=U(x_2)$ is a parallel base flow directed along the $x_1$ axis and
depending on the $x_2$ (wall-normal) coordinate only, $u_i$ denote perturbation
velocity components and $p$ the perturbation pressure, $\rho$ and $\mu$ the constant density
and viscosity of the incompressible fluid, and the Einstein summation convention
is implied on repeated mute indices. $s_{ij}$ are the components of a
random white-noise stress tensor whose correlation function $S_{ijhk}=\left< s_{ij}(t,\vec{x})s_{hk}(t',\vec{x}')\right>$ was given in \cite{LL} as
\begin{equation}
S_{ijhk} = 2\mu k_B T\left(\delta_{ih}\delta_{jk} +
\delta_{ik}\delta_{jh}\right)\delta(t-t')\delta(\vec{x}-\vec{x}')
\label{Landau}
\end{equation}
with $k_B$ denoting the Boltzmann constant and $T$ the (unperturbed) temperature.

As is customary in parallel-flow stability theory, eqs.(\ref{physdom}) can be
Fourier-transformed with respect to time and to the homogeneous directions $x_1$ and $x_3$, and thus reduced to ordinary differential
equations in the only independent variable $x_2$ which we shall henceforth also denote as $y$. We shall follow the convention $u_i\sim \exp\left(\im \omega
t - \im \alpha x_1 - \im \beta x_3\right)$. Therefore in the following the derivative $\partial/\partial t$ is understood as a multiplication by $\im\omega$, the derivative $\partial/\partial x_1$ as a multiplication by $-\im\alpha$ and the derivative $\partial/\partial x_3$ by $-\im\beta$, although where convenient for tensorial notation we shall continue to denote them as derivatives.

\section{Adjoint equations}
The homogeneous version of eqs.(\ref{transformed}), in which they are most frequently
encountered, is an eigenvalue problem. Nontrivial solutions are found when either
$\omega$ (temporal stability problem) or $\alpha$ (spatial stability problem) is
allowed complex values, \ie\ more rigorously when either the time- or the
$x_1$- Fourier transform is replaced by a Laplace transform. To study the inhomogeneous problem (\ref{transformed}), on the other hand, we can
obtain a substantial simplification by the use of adjoint equations. The adjoint problem is introduced by
premultiplying the homogeneous form of eq.(\ref{cont}) by a new variable $c^+$
(continuity adjoint) and the homogeneous form of
eq.(\ref{mom}) by a new variable $q^+_{j}$ (momentum
adjoint) \footnote{The superscript $^+$ does not stand for complex conjugate or
conjugate transpose as is customary in the context of self-adjoint problems,
but for a completely new variable that obeys its own set of differential
equations and represents a sensitivity to the \rhs\ of respectively the
continuity and momentum balance equations. To avoid confusion we refrain from naming these variables adjoint velocity and
adjoint pressure as was sometimes done by past authors.} and integrating
over the entire $y$-domain to obtain the so called Lagrange identity:
\begin{multline}
\int \left\{
 c^+\left(\de{u_i}{x_i}\right) +q^+_j\left[\rho\left(\im\omega u_j - \im\alpha U u_j + \delta_{1j}\frac{dU}{dx_2}u_2\right)\right.\right.\\
 \left.\left. +\de{p}{x_j} -
\mu \de{^2u_j}{x_i\partial x_i}  - \de{s_{ij}}{x_i}\right]
\right\} \d y = 0\, .
\label{LagrId}
\end{multline}
The Lagrange identity must be satisfied for whatever \rhs\ of
eqs.(\ref{transformed}), \ie\ for whatever $u_i$ and $p$ that obey the boundary
conditions but not the (homogeneous) direct equations. This is imposed by
transforming the integral through integration by parts 
until all derivatives of $u_i$ and $p$ disappear; the coefficients of
these variables in the integral are then set to zero, and there results an eigenvalue problem for the unknowns $c^+$ and $q^+_{i}$:
\begin{multline}
\de{q^+_i}{x_i} = 0\, ,\quad
\rho\left(\im\omega q^+_j - \im\alpha U q^+_j + \delta_{2j}\frac{dU}{dx_2}q^+_1\right)\\
-\de{c^+}{x_j} - \mu \de{^2q^+_j}{x_i\partial x_i}=0 \, .
\label{adjoint}
\end{multline}
Integration by parts at the same time provides the appropriate boundary
conditions for these equations.

Here we consider the temporal stability problem (simpler than, but analogous to, the practically more interesting spatial stability problem). If eqs.(\ref{adjoint}) are solved for a specific (generally complex)
eigenvalue $\omega^{(1)}$, whereas in eqs.(\ref{transformed}) $\partial/\partial t$ is replaced
by the Laplace variable $\sigma$, premultiplying eqs.(\ref{transformed}) by
respectively $c^{+(1)}$ and $q^{+(1)}_j$ and subtracting (\ref{LagrId}) gives
\begin{multline}
(\sigma - \im\omega^{(1)}) \rho\int q^{+(1)}_j u_j \d y = \int q^{+(1)}_j \de{s_{ij}}{x_i}  \d y =\\
= \int  - \de{q^{+(1)}_{j}}{x_i} s_{ij} \d y\, ,
\label{coeff1}
\end{multline}
the last step being once again an integration by parts. Since direct and adjoint modes are mutually orthonormal with respect to the
scalar product $\int q^{+}_j u_j \d y$, the \lhs\ of eq.(\ref{coeff1}) contains the
complex amplitude $C^{(1)}$ of the mode $u_j^{(1)}$ of frequency $\omega^{(1)}$
in the modal representation of $u_i$:
\begin{equation*}
C^{(1)} = \int q^{+(1)}_j u_j \d y = \rho^{-1} \left(\sigma - \im\omega^{(1)}\right)^{-1}\int  R_{ij} s_{ij} \d y\, ,
\end{equation*}
where $R_{ij}= - \partial q^{+(1)}_{j}/\partial x_i$. In the time domain
\begin{equation}
C^{(1)} = \rho^{-1} \int_0^t\e^{\im\omega^{(1)}(t-\tau)}\int R_{ij} s_{ij} \d y  \;\d \tau\, .
\label{coeff}
\end{equation}

\section{Stochastic disturbances}
The formalism of the previous sections provides the response of a single mode of our
shear flow to a stress forcing of arbitrary, deterministic or random, nature. Armed
with these results we can now go back to analyse the random stresses that arise
from thermal fluctuations.
Since $s_{ij}$ is a random white noise, of which only the correlation function
is known, $C^{(1)}$ will be a stochastic variable too. The mean square of its
value is given by
\begin{multline}
\left<\left|C^{(1)}\right|^2\right> 
= \rho^{-2}\iint\left<\left(\iint \e^{\im\omega^{(1)}(t-\tau)}  R_{ij} s_{ij} \d y  \,\d
\tau\right)^*\right.\\
\left.\times \left(\iint \e^{\im\omega^{(1)}(t-\tau')} R_{hk}
s_{hk} \d y'  \,\d \tau' \right) \right>
\frac{\d\alpha}{2\pi}\,\frac{\d\beta}{2\pi}=\\
= \rho^{-2}\iint\!\!\!\iiiint
\e^{\im\omega^{(1)}(t-\tau')-\im\omega^{(1)*}(t-\tau)}\\
\times R_{ij}^*(y)
R_{hk}(y')S_{ijhk}\d y  \,\d\tau \,\d y'  \,\d
\tau'\, \frac{\d\alpha}{2\pi}\,\frac{\d\beta}{2\pi}\, .
\end{multline}
On account of eq.(\ref{Landau}) two integrals are eliminated by
corresponding $\delta$-functions, whereas the exponential can be explicitly
integrated in time. The final result is
\begin{multline}
\left<\left|C^{(1)}\right|^2\right> = \frac{\mu k_B
T}{\rho^2} \iint\left[\frac{1 -
\e^{-2\Im(\omega^{(1)})t}}{2\Im(\omega^{(1)})}\right.\\
\times \left. \int(R_{ij}+R_{ji})^*(R_{ij}+R_{ji})\, \d y \right] \frac{\d\alpha}{2\pi}\,\frac{\d\beta}{2\pi}\, .
\label{msv}
\end{multline}

\section{Discussion and order-of-magnitude considerations}
When the mode with eigenvalue $\omega^{(1)}$ is stable ($\Im(\omega^{(1)})>0$), the \rhs\
of eq.(\ref{msv}) converges to a steady state for $t\rightarrow\infty$, and the
fluctuations induced by thermal noise have a finite limiting amplitude. When
this mode is unstable, on the other hand, the perturbation diverges in time and
no statistically steady state is attained; if the instability has a well-defined
origin in time (thermal noise cannot be switched on and off, of course, but the
base flow can), eq.(\ref{msv}) allows us to determine the duration of the
transient until the perturbation attains an amplitude comparable to the base
flow and transition to turbulence occurs. Even more interestingly, an analogous
process occurs in space: a spatial-stability analysis (the details of which will be given in a separate paper) allows us to
determine the length of the region where laminar flow prevails until the
perturbation attains an amplitude comparable to the base 
flow and transition to turbulence occurs.

Before proceeding with a numerical calculation, a dimensional analysis may be useful. The dimensions of $q^{+(1)}_i$ are those of (velocity$\times$length)$^{-1}$, \ie\ TL$^{-2}$ if we use T and L to denote time and length measuring units. The dimensions of $R_{ij}$ are thus TL$^{-3}$, and since those of $(\mu k_B T)/\rho^2$ are T$^{-3}$L$^7$, the \rhs\ of eq.(\ref{msv}) is dimensionless as it should be. If all lengths are normalized with a reference length $h$ (a representative breadth of the shear flow, such as a boundary-layer thickness or channel
height) and all velocities with a reference velocity $V$ (representative of
the base velocity profile, such as its maximum value), viscosity becomes
encapsulated in the Reynolds number $Re=hV/\nu$ (where as usual $\nu=\mu/\rho$);
the coefficient $(\mu k_B T)/\rho^2$ gives rise to a new dimensionless number
that we shall conveniently define as the ratio of
two lengths $\lambda/h$, where:
\begin{equation}
\lambda = k_B T/\rho\nu^2\, .
\label{LL}
\end{equation}
The so defined characteristic length $\lambda$ is a material property; for standard air at 300~K and atmospheric
pressure, for instance, $\lambda=1.508\times 10^{-11}$ m. Equation (\ref{msv}) may thus be rewritten in dimensionless form as
\begin{multline}
\left<\left|C^{(1)}\right|^2\right> = \frac{\lambda}{h} Re^{-3}
\iint\left[\frac{1 - \e^{-2\Im(\omega^{(1)})t}}{2\Im(\omega^{(1)})}\right.\\
\times \left. \int(R_{ij}+R_{ji})^*(R_{ij}+R_{ji})\, \d y \right] \frac{\d\alpha}{2\pi}\,\frac{\d\beta}{2\pi}\, ,
\label{ndmsv}
\end{multline}
where all quantities inside the integral are nondimensionalized with the
reference length $h$ and velocity $V$.

An obvious use of eq.(\ref{ndmsv}) is to make a transition
prediction by comparing the value of $C^{(1)}_{\mrms}$ to
the level of disturbances that is experimentally observed to occur right before the breakdown to
turbulence. An order-of-magnitude estimate of the expected threshold
can be obtained through the following considerations: the denominator $\Im(\omega^{(1)})$ is
a small number almost comparable with $Re^{-1}$ (the exponent is not quite $-1$, according to
the asymptotic theory of the Orr-Sommerfeld equation, but close enough for the
present estimate); the threshold \rms\ velocity perturbation is of the order of
$1\div 2\%$ \cite{K} while the order of magnitude of $R_{ij}$ is typically larger than unity because of the non-normality \cite{TTRD} of fluid-dynamic stability problems. If we assume that all together these factors (squared) compensate the remaining factor of $Re^{-2}$, we can roughly equal the exponential $\e^{-2\Im(\omega^{(1)})t}$, which would be $e^{2N}$ according to the $e^N$ method, to $h/\lambda$. For $h=1$~mm (corresponding to $\delta_{99}\approx 5$~mm or $L\approx 1.9$~m) in air, we thus obtain
\[
N=0.5\log(h/\lambda)=9.00 \,.
\]

Whereas this simple estimate cannot, of course, be trusted to three decimal digits, its value is strikingly on top of the empirical value of $9\div 10$ neper used in aerodynamics. This is how we got to realize that receptivity to microscopic fluctuations could have a significant role in causing transition to turbulence. Owing to the presence of the logarithm, changing thickness ($h$) or type of fluid ($\lambda$) just moderately affects this estimate.

\begin{figure}
\vspace{0.4cm}
\hspace*{-0.8cm}\input{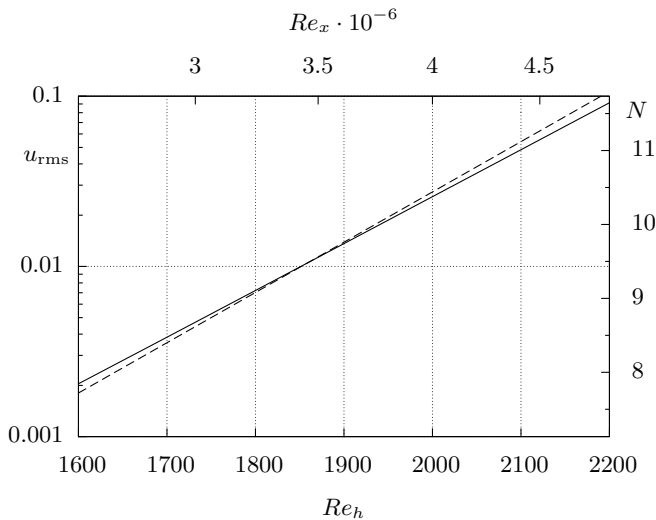}
\vspace{-0.7cm}
\caption{Longitudinal-velocity fluctuation $u_\mrms$ \textit{vs.} local Reynolds number, as
obtained from a numerical integration of the spatial receptivity problem in a
Blasius boundary layer (solid), compared with the amplification envelope of the
standard $e^N$ method (dashed). $u_\mrms$ attains $1$\% of the freestream velocity in the same region where $N\approx 9\div 10$.}
\label{fig:1}
\end{figure}

A precise numerical evaluation of the correspondent of eq.(\ref{ndmsv}) for the quasi-parallel case of engineering interest where
amplification occurs in space rather than in time is the subject of a separate,
longer paper. Preliminary numerical results in Figure \ref{fig:1} 
quantitatively confirm that even in the absence of remnant atmospheric
sound or vorticity (considered until today to be indispensable), transition to
turbulence on airplane wings would occur where it is practically observed
to occur. For, microscopic fluctuations of thermal origin set a finite and
impassable lower threshold to the input noise of this like any other amplifier.

\end{document}